\documentclass[article,reprint,superscriptaddress]{revtex4-2} 
\usepackage{graphicx}
\usepackage{amsmath}
\usepackage{mwe}
\usepackage{stackengine}
\usepackage[caption=false]{subfig}
\usepackage{dcolumn}
\usepackage{siunitx}

\usepackage[usenames,dvipsnames,svgnames,table]{xcolor}
\usepackage[normalem]{ulem}
\newcommand{\gd}[1]{\textcolor{black}{#1}}
\newcommand{\fp}[1]{\textcolor{black}{#1}}

\graphicspath{ {./images/} }

\begin{document}
\title{Quantum correlations, mixed states and  bistability at the onset of lasing}

\author{Francesco Papoff}
\email{f.papoff@strath.ac.uk}
\affiliation{Department of Physics, University
of Strathclyde,  107 Rottenrow,  Glasgow G4 0NG, UK.}
\author{ Mark Anthony Carroll}
\affiliation{Department of Physics, University
of Strathclyde,  107 Rottenrow,  Glasgow G4 0NG, UK.}
\author{Gian Luca Lippi}
\affiliation{Universit\'e C\^ote d'Azur, Institut de Physique de Nice, UMR 7710 CNRS, 17, rue Julien Laupr\^etre, 06200 Nice, France}
\author{Gian-Luca Oppo}\affiliation{Department of Physics, University
of Strathclyde,  107 Rottenrow,  Glasgow G4 0NG, UK.}
\author{Giampaolo D'Alessandro}
\affiliation{School of Mathematics, University of Southampton, Southampton SO17 1BJ, United Kingdom}

\date{\today}

\begin{abstract} 
  We derive a model for a single mode laser that includes all two particle quantum correlations between photons and electrons. In contrast to the predictions of semi-classical models, we find that lasing takes place in the presence of quantum bistability between a non-lasing and a non-classical coherent state.  The coherent state is characterized by a central frequency and a finite linewidth and emerges with finite amplitude from a saddle-node bifurcation together with an unstable coherent state.  Hence coherent emission in nanolasers  originates through a mixing of lasing and non-lasing states. In the limit of a macrolaser with a large number of emitters and non-resonant modes, the laser threshold approaches the prediction of the semi-classical theory, but with the important difference that lasing can be achieved only in the presence of finite size perturbations. 
  \end{abstract}

\maketitle

The development of nanolasers in recent years~\cite{tempel11a,koulas-simos23a} has been driven by the demand for devices with minimal footprint and thermal load for applications such as on-chip communications, sensing, and biological probes~\cite{ma2019applications}, with potential non-classical applications enabled by photon number squeezing~\cite{Kreinberg2017a,mork22a,carroll21b}.  The size of these devices raises interesting fundamental questions about the identification of the lasing threshold~\cite{ning2013laser, carroll21a, carroll21b} and the role of light-matter quantum correlations. These are expected to affect the dynamics and the statistics of light emission in nanolasers more strongly than in macroscopic lasers, where the large number of cavity modes and intracavity emitters ensure the validity of the semi-classical limit~\cite{Narducci1988}.  The search for performing models is of paramount importance at the nanoscale, where the challenging nature of the signals (low photon flux and large bandwidth) restricts the experimental information to the photon statistics.  Reliable predictions become the only source for a meaningful comparison and interpretation of experimental results. For small numbers of emitters and electromagnetic field modes, laser quantum models become crucial under the requirement of overlapping their predictions with those of semi-classical theories when reaching the macroscopic scale.

Quantum models for nanolasers have been developed by using Heisenberg-Langevin equations~\cite{protsenko22a}, density matrix theory~\cite{yacomotti23a}, non-equilibrium Green's functions~\cite{vyshnevyy22a} or by applying the cluster expansion~\cite{Fricke1996}, where the fast variables associated with coherent emission are neglected~\cite{chow13a,chow14a, Kreinberg2017a, gies07a} and only the slowly varying quantum correlations are kept. These models have recently been extended within a semi-classical theory, by including the expectation values of the coherent field and the medium polarization, and by neglecting electron-electron and fast photon-electron quantum correlations \gd{(Coherent-Incoherent Model, CIM)}~\cite{carroll21a,carroll21b}.  The introduction of coherent variables has allowed us to identify a laser threshold, which can be experimentally detected by measuring the first-order correlation $g^{(1)}(\tau)$~\cite{tempel11a}, beyond which stimulated emission becomes continuous~\cite{carroll23a}.

In this paper we address two fundamental questions: to what extent do quantum correlations affect lasing, and whether models that include them can reduce to the semi-classical theory as the number of intracavity emitters grows. We consider all two-particle quantum correlations (photon-photon, photon-electron, and electron-electron) in a nanolaser model~\cite{Kreinberg2017a} containing quantum dots at cryogenic temperature \gd{(Two-Particle Model, TPM)}. The quantum dots have two localized levels where electrons and holes are injected from the wetting layer, and the light-matter coupling is weak.  Coulomb~\cite{gies07a, chow13a} and phonon~\cite{baer06a} scattering are considered through dephasing terms and the interaction Hamiltonian is simplified by keeping only \gd{terms that do not oscillate at the light frequency scale} (rotating wave approximation in the weak coupling regime)~\cite{carroll21a,carroll21b}. The bosonic operators $b^\dagger, b$ describe photon creation and annihilation processes, and the fermionic operators $c^\dagger, c$ ($v^\dagger, v$) represent the upper (lower) energy level electron creation and annihilation.  To understand the physical meaning of the theory, it is important to associate the operators with particles: $b^\dagger, b$ are single-particle bosonic operators, while the single-particle fermionic operators~\cite{kira2011semiconductor} are composed of pairs of operators such as the electron number in the upper-level, $c^\dagger c$, and the polarization, $c^\dagger v$.

The interaction between light and matter gives rise  to an infinite hierarchy of differential equations for the expectation values of operators involving an increasing number of particles. The hierarchy can be truncated to a finite set by noting that the expectation value of any $M$-particle operator is the sum of an $M$-particle correlation -- originated by processes involving all the $M$ particles -- and products of expectation values of operators with numbers of particles ranging from $1$ to $(M-1)$~\cite{kira2011semiconductor}. 

\fp{A detailed analysis of different approximation schemes has been performed in \cite{sanchez-barquilla20a}, where it was shown that some approximations can lead to non physical results such as negative populations of the excited state. Our model considers only two-particle correlations, which, in the weak coupling regime, are expected to be larger than higher order correlations \cite{Fricke1996} and we have verified that it produces physical results for both populations and photon numbers.}

\gd{To implement this scheme } we use the identity $\langle O_i O_j O_k \rangle = \langle O_i O_j O_k \rangle_\mathcal{C} + \langle  O_i O_j O_k  \rangle_\mathcal{D}$, where $O_i,O_j,O_k$ are arbitrary one particle operators, $\langle O_i O_j O_k \rangle_\mathcal{C}$ is the three particle correlation and $\langle O_i O_j O_k  \rangle_\mathcal{D}$ the sum of products of single and two particle expectation values. We then truncate the equations at two particle level by setting $\langle O_i O_j O_k \rangle_\mathcal{C} = 0$.  We apply this procedure to the exact (not-closed) equation
\begin{equation}
  \label{eq:3}
   \frac{\mathrm{d}}{\mathrm{d}t}
  \begin{bmatrix}
    \langle O_i  \rangle  \\  \langle  O_i {O}_j  \rangle
  \end{bmatrix}
  = L \begin{bmatrix}
    \langle O_i \rangle  \\  \langle  O_i O_j \rangle
  \end{bmatrix}
  +  \begin{bmatrix}
\langle \mathcal{A}( O_i ) \rangle \\ \langle \mathcal{A}( O_i O_j ) \rangle
  \end{bmatrix}
  + \begin{bmatrix} 0  \\ R \langle O_i O_j O_k \rangle \end{bmatrix}
\end{equation}
where $L$ is a square matrix,  $\mathcal{A}$  accounts for  the presence of non resonant modes, Eqs.~(S.2-3) in the Supplementary Material, and $R$ is a rectangular matrix that describes the coupling between $2$ and $3$ particle expectation values. The end result is an approximate (closed) system of equations obtained by replacing the last two terms in equation~\eqref{eq:3} by
\begin{subequations}
  \label{eq:4}
  \begin{align}
    \label{eq:2}
    \begin{bmatrix}
      \langle \mathcal{A}( O_i ) \rangle \\
      \langle \mathcal{A}( O_i O_j ) \rangle
    \end{bmatrix}
    & \longrightarrow
      \begin{bmatrix}
        \tilde{\mathcal{A}}( \langle O_i \rangle ) \\
        \tilde{\mathcal{A}}( \langle O_i \rangle \langle O_j \rangle) 
      \end{bmatrix} ,\\
    \label{eq:5}
    \begin{bmatrix} 0  \\ R \langle O_i O_j O_k \rangle \end{bmatrix}
    & \longrightarrow
      \begin{bmatrix}
        0  \\ R  \langle  O_i O_j O_k  \rangle_\mathcal{D}
      \end{bmatrix} , 
  \end{align}
\end{subequations}
where $\tilde{\mathcal{A}}$ are functions that give the radiative decay and pumping terms resulting from the adiabatic elimination \fp{of the correlations of the non resonant modes with the polarization, see Eq.~(12) in \cite{gies07a} and Eqs.~(S.10-11)}. While equation~\eqref{eq:3} is linear, the terms $\tilde{\mathcal{A}}( \langle O_i \rangle \langle O_j \rangle)$ and $R \langle O_i O_j O_k \rangle_\mathcal{D}$ are nonlinear.  More generally, we can truncate the equations at any $M$ number of particles by neglecting all correlations of $M+1$ particle operators~\cite{leymann2013expectation,leymann14a}, and thus approximate the eigenstates of a linear open quantum system of very large (possibly infinite) dimension with the stable steady states of a nonlinear system of much smaller dimension.  For simplicity sake, we consider here a set of $N$ identical single--electron quantum dots~\cite{Simmons2007}.  In this case, Eqs.~\eqref{eq:3}-\eqref{eq:4} are equivalent to a set of 12 complex equations, \gd{the TPM equations with identical quantum dots,}\gd{Eqs.~(S.12) }. 


We determine the role of different two particle correlations by comparing \gd{the TPM and CIM predictions~\cite{carroll21a, carroll23a}.} The CIM variables comprise three slow variables, electron number $\langle c^\dagger c\rangle$, photon-assisted polarization $\langle b c^\dagger v\rangle$, and photon number $\langle b^\dagger b \rangle$, and two complex fast variables oscillating at the laser frequency: coherent field amplitude $\langle b \rangle$ and polarization $\langle v^\dagger c\rangle$. The TPM has additional photon-electron and photon-photon fast variables including two photon-electron expectation values, namely $\langle b c^\dagger c \rangle$, and $\langle b v^\dagger c \rangle$, and one fast photon-photon expectation value $\langle b b \rangle$. Additionally, it has electron-electron expectation values comprising two slow variables, $\langle c^\dagger v^\dagger c v\rangle$, $\langle c^\dagger c^\dagger c c\rangle$, and two fast variables, $\langle v^\dagger c^\dagger c c\rangle$, $\langle v^\dagger v^\dagger c c\rangle$. We model phonon scattering by adding a dephasing term to electron-electron expectation values, $\mu \gamma$, with $\mu \ge 0$, following Ref.~\cite{baer06a}, where it was shown that this dephasing reproduces the key effects of a microscopic theory of phonon scattering. The other control parameters are common to all models: the light-matter coupling $g$; the decay rates of the upper-level population due to non-radiative processes, $\gamma_{nr}$, and non-lasing modes, $\gamma_{nl}$; the dephasing rate of the polarization, $\gamma$; the pump rate per emitter, $r$; and the number of quantum dots $N$ inside the laser. \gd{All the parameter values have been chosen based on experimental considerations, see section~S-VII. }
\gd{For both CIM and TPM we characterise the lasing solution using its coherence time, which can be experimentally measured through the visibility of interference fringes~\cite{tempel11a}. The richer non-linear structure of the TPM with respect to the CIM equations is expected to lead to multistability, which, as shown in similar situations~\cite{ivanchenko17a,bartolo16a,casteels17a}, corresponds to mixed states in the open quantum system.}

\begin{figure}
  \centering
  \begin{tabular}{c}
    \includegraphics[width=0.45\textwidth]{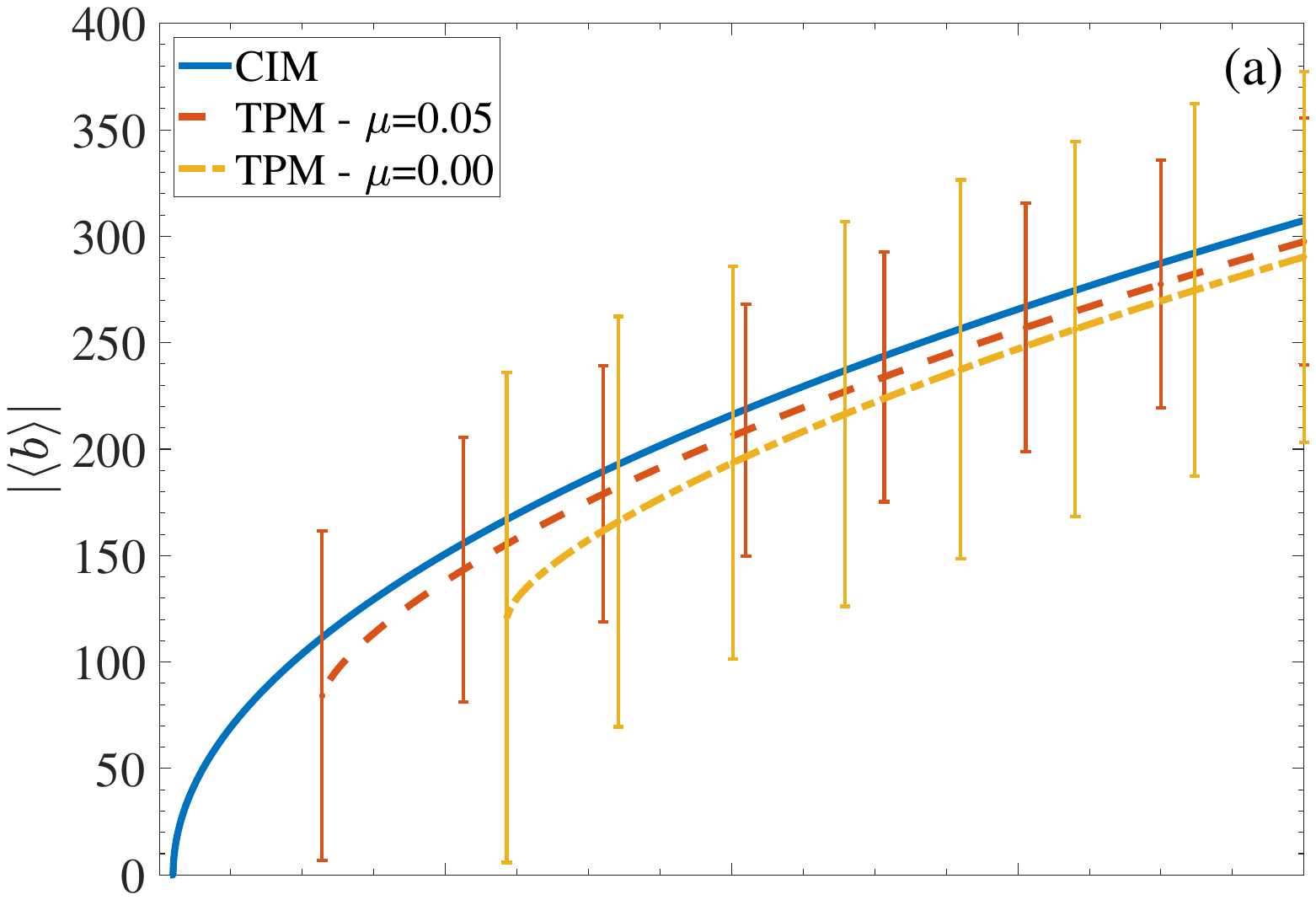}\\*[-6.5mm]
    \includegraphics[width=0.45\textwidth]{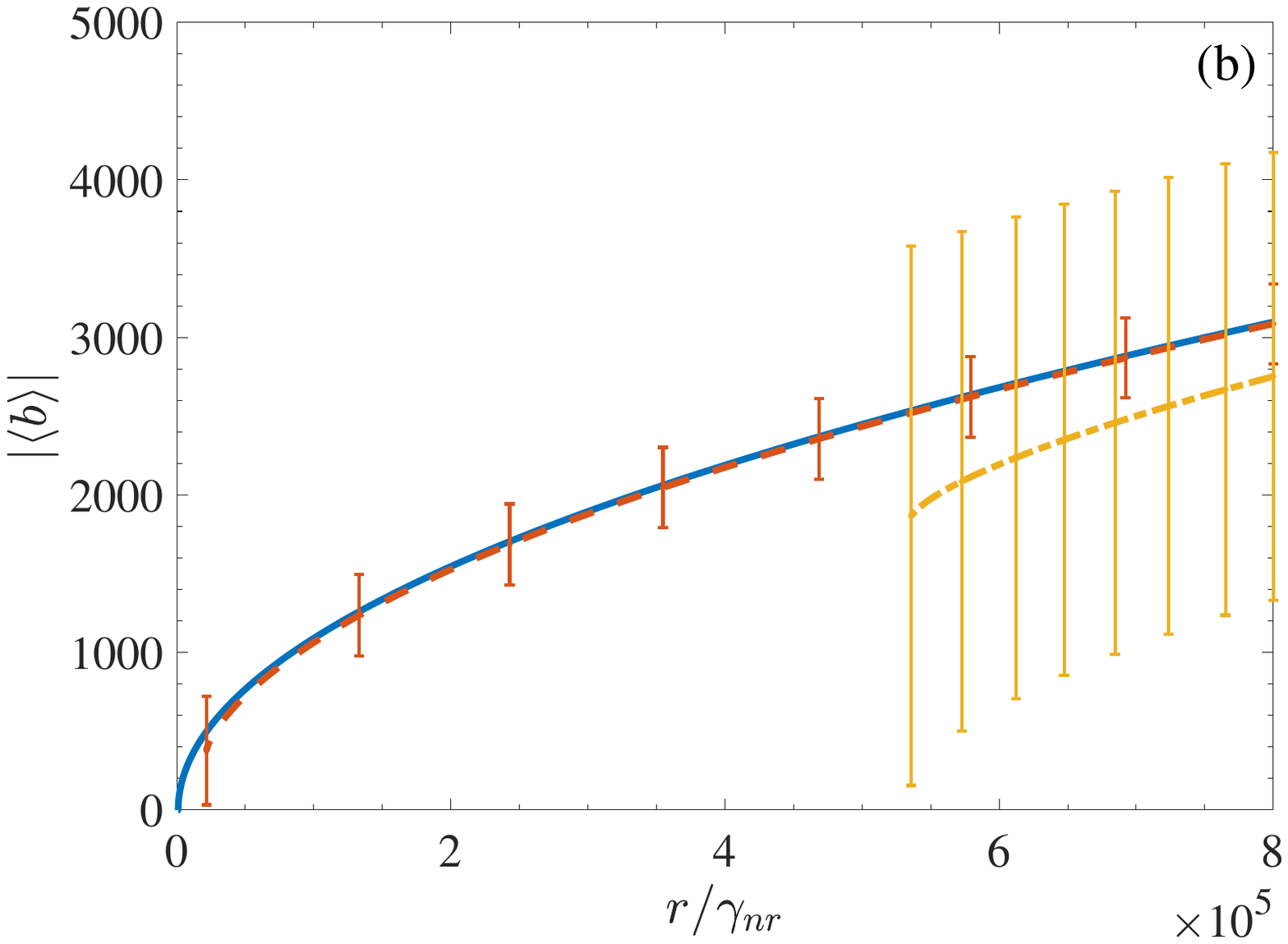}
  \end{tabular}
  \caption{Lasing solutions for the CIM (solid blue line) and the TPM (dashed lines, yellow for $\mu=0$, red for $\mu=0.05$) with $N=25$ (a) and $N=500$ (b). The thin vertical lines measure the generalized standard deviation $\sqrt{|\langle b b \rangle - \langle b \rangle^2|}$.  Time, decay rates and coupling parameters are scaled with $\gamma_{nr}=10^{-9}\si{s^{-1}}$. Other parameter values (common to all figures) are $g=70$, $\Delta \nu=0$, $\gamma = 10^{4}$, $\gamma_{c} = 10$, and $\gamma_{nl}=0$ \gd{(equivalent to $\beta=1$)}.}
  \label{fig:BD}
\end{figure}

In Fig.~\ref{fig:BD} we show the absolute value of the amplitude of the coherent field, $|\langle b \rangle|$, versus the ratio of the pump rate per emitter and the non-radiative decay rate, $r/\gamma_{nr}$, for nanolasers with $N=25$ (Fig.~\ref{fig:BD}a) and $N=500$ (Fig.~\ref{fig:BD}b)\gd{, computed by integrating Eqs.~(S.14).} The vertical bars, of length equal to the generalized standard deviation $\sqrt{|\langle b b \rangle - \langle b \rangle^2|}$, illustrate both the variation of the values of $|\langle b \rangle|$ and 
that this state is not a Glauber coherent state --  the analog of a coherent classical field with with $\langle b b \rangle - \langle b \rangle^2=0$ -- but a non-classical Gaussian state~\cite{olivares12a}. The key result is that for the TPM the coherent field has always a finite threshold amplitude, while this is zero in the semi-classical CIM (solid, blue line).

Electron-electron correlations increase both the thresholds and the generalized standard deviation even for a large number of QDs.  These effects, however, are severely reduced by the presence of a small amount of phonon scattering\gd{, e.g. the $\mu =0.05$ curves in Fig.~\ref{fig:BD}}. This sensitivity is in agreement with the results of~\cite{baer06a,carroll22a}. In the presence of phonon scattering, the amplitude of the coherent field approaches that of the semi-classical theory as the number of emitters increases. A similar dependence  on quantum correlations of the dynamics of large numbers of particles has been recently found also in spin systems~\cite{fowler-wright23a}. 

\begin{figure}
\centering
  \includegraphics[width=0.45\textwidth]{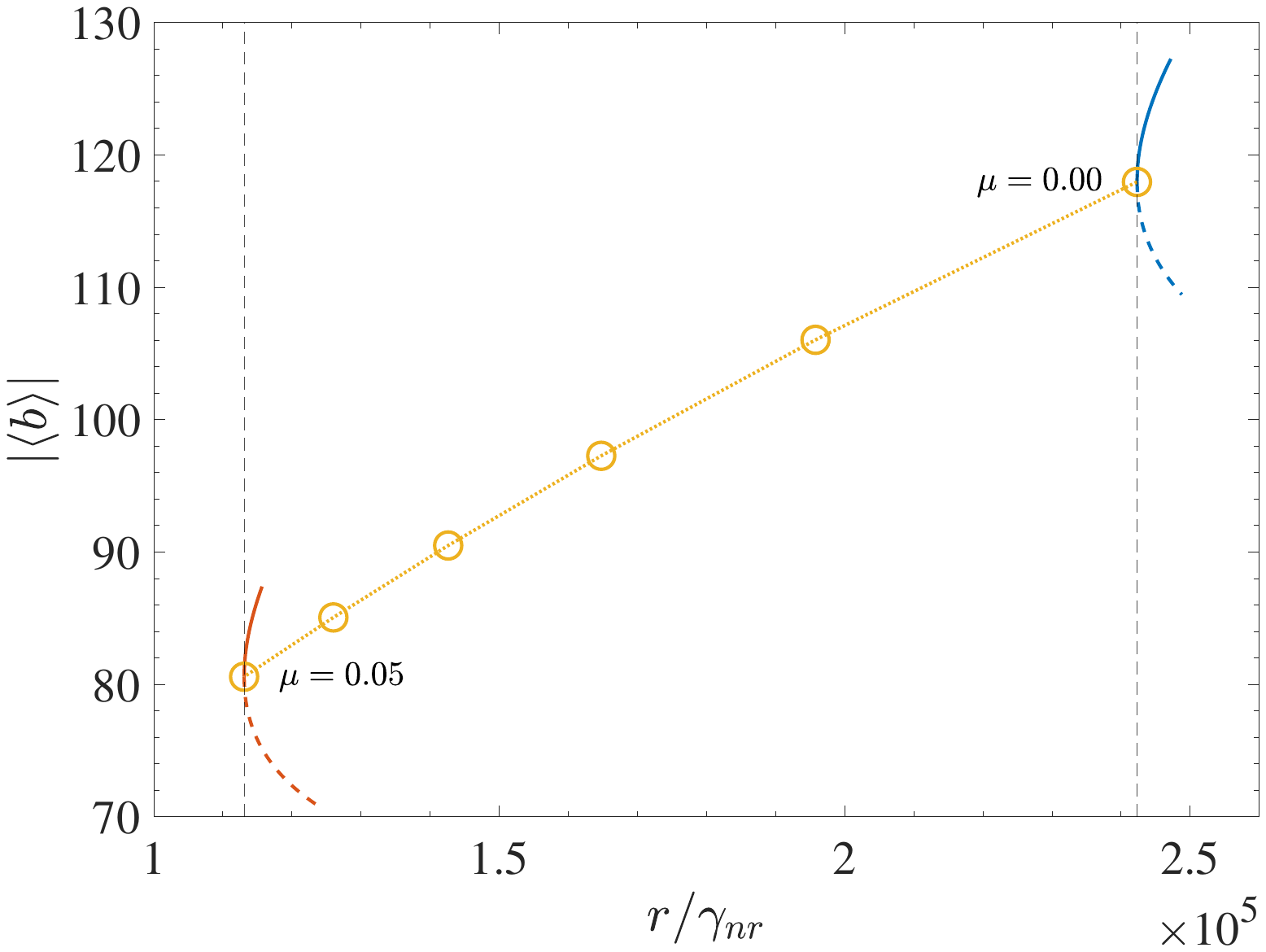}
  \caption{Bifurcation diagram of the \gd{TPM} lasing solution (stable, solid, unstable, dashed) for $N=25$ as a function of the pump for $\mu=0.00$ (blue lines) and $\mu=0.05$, (red lines). The bifurcation points where the stable and unstable solutions meet are the lasing thresholds.  They are indicated by open circles for equally spaced values of $\mu$ in the range $[0,0.05]$. \fp{The incoherent solution $|\langle b \rangle| = 0$ \gd{(not shown) is stable. }} The dotted line is only a guide for the eye. The vertical dashed lines show the analytical values of the threshold provided by Eq.~\eqref{r_tpm}.}
  \label{fig:saddle-node}
\end{figure}

Furthermore, contrary to what happens in the semi-classical theory, there are two lasing solutions, one stable and one unstable, that appear at the lasing threshold through a \gd{Hopf} saddle-node bifurcation, Fig.~\ref{fig:saddle-node}\gd{, which can be seen as an \fp{extension} of the standard CIM lasing Hopf bifurcation~\cite{carroll21a, Golubitsky1981Classification} }.  \fp{The bifurcation appears when the correlation between the population of the upper level and the field -- a non-classical correlation emerging from the interaction between field and quantum dots -- is introduced, even if all other correlations are neglected. There are two observable stable states, one with coherent emission and the other without, and an unstable state. Physically, the unstable state is a state with coherent emission and extremely short life- time.}
An accurate analytical estimate of the laser threshold (vertical dashed lines in Fig.~\ref{fig:saddle-node}) is found from the analytical solution of the incoherent state,
\begin{equation}
  \label{r_tpm}
  r_{th}= \left( \Gamma_n + \frac{\Gamma_g}{K} \right)
  \frac{\langle c^\dagger c \rangle_{th}}{1-\langle c^\dagger c \rangle_{th}},
\end{equation}
where $\langle c^\dagger c \rangle_{th}$ is the CIM threshold value, Eq.~(S.29), $\Gamma_n = \gamma_{nr} +\gamma_{nl}$, $\Gamma_g = 2 g^2 (\gamma + \gamma_c)/[(\gamma + \gamma_c)^2 + \Delta \nu^2]$ and
$K=1-\Gamma_g (2 \langle c^\dagger c \rangle_{th} - 1) \{N/2 \gamma_c + (N-1)/[2 \gamma (1+\mu)]\}$.

\fp{For completeness, we note there are other ways to define the threshold~\cite{saldutti24a}, which appear to identify different points in the transition from incoherent to coherent emission in nanolasers and behave differently from very large values of $g$. Eq.~\eqref{r_tpm} and Eq.~(S.29) identify the appearance of \gd{a} stable coherent \gd{solution } with constant amplitude; based on ~\cite{habib23a} we expect that below \gd{this threshold, transients of coherent emission are possible and can lead to a gradual increase of time averaged measures of the coherence as the pump approaches the threshold value. }}  Moreover, the non-lasing solution continues to be linearly stable above the lasing threshold. It should be noted that this bistability does not appear in the semi-classical limit, where, in a framework where quantum correlations vanish, bistability is associated with a first-order phase transition. Here, however, bistability is a consequence of the inclusion of two-particle quantum correlations and indicates the presence of a mixing of lasing and non-lasing states in nanolasers far from the semi-classical limit.  In Fig.~\ref{fig:saddle-node} we also plot the effect of phonon scattering on the lasing threshold. For all values of $\mu$ plotted there are a stable and an unstable lasing solution at threshold.  This implies that the bistability is present also when the laser threshold and the $|\langle b \rangle|$ curve approach the semi-classical theory, showing that lasing can be started only by finite amplitude perturbations.

\begin{figure}
  \centering
  \begin{tabular}{c}
    \includegraphics[width=0.45\textwidth]{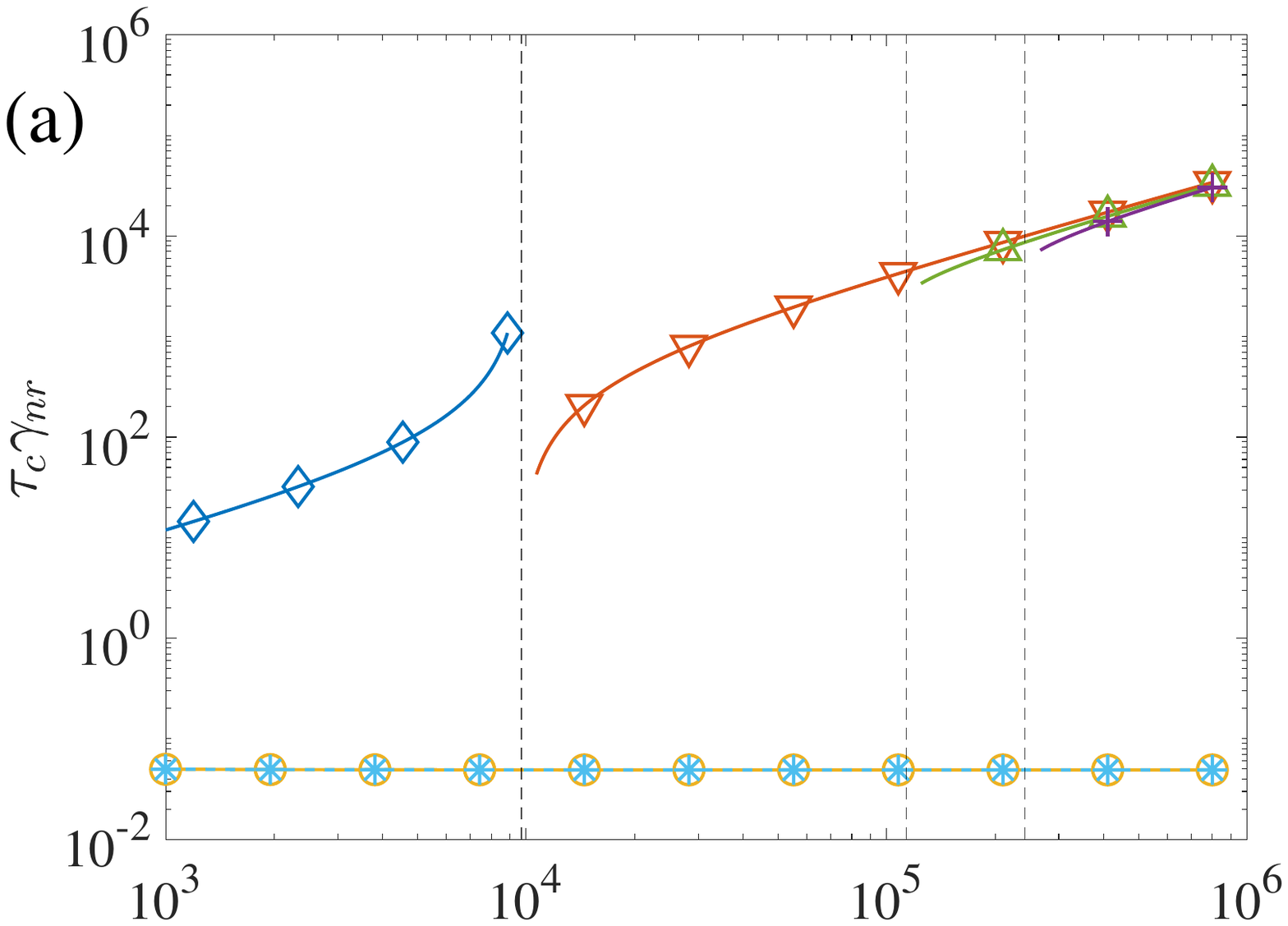}\\*[-2.5mm]
    \includegraphics[width=0.45\textwidth]{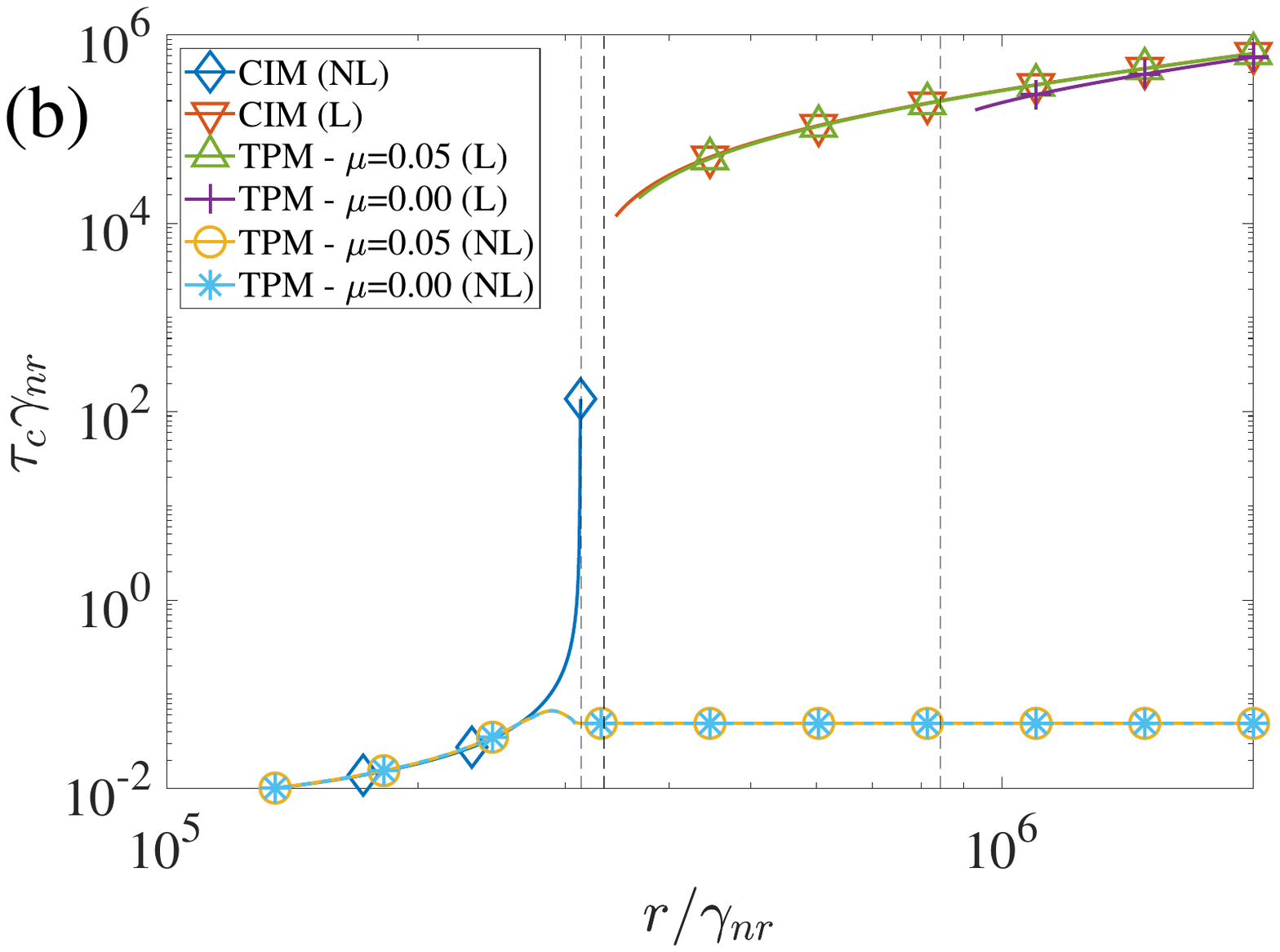}
  \end{tabular}
  \caption{Decay time of the first order correlation function, $\tau_{c}$, for lasing (L) and non-lasing (NL) CIM and TPM solutions \fp{for a nanolaser (a) and a macrolaser (b)}. The vertical dashed lines show from left to right the analytical values of the threshold, Eq.~\eqref{r_tpm}, for the CIM, TPM $\mu=0.05$ and TPM $\mu=0.00$ lasing solutions. The correlation time of the CIM~(NL) solution diverge at threshold. The CIM~(NL) and TPM~(NL) curves were computed using Eqs.~(S.27) and~(S.24), respectively. The TPM~(L) curves were computed using the Schawlow-Townes theory, Eq.~(S.25), and plotted only for pump values at least 10\% above threshold. The parameters are \fp{$\beta=1$ and $N=25$ for the nanolaser in (a), $\beta=3.4\times 10^{-6}$ and $N=500$ for the macro laser in (b)}. All other parameters as in Fig.~\ref{fig:BD}. 
  }
  \label{fig:g1}
\end{figure}

To illustrate further the difference between the CIM and TPM lasing and non-lasing solutions \fp{and the impact these have on \gd{nano- and macrolasers}, we plot their first order coherence time, $\tau_{c}$, defined in Eq.~(S.17), \gd{for a nano- and  a macrolaser, Fig.~\ref{fig:g1}(a) and~(b) respectively}}. For all non-lasing solutions $g^{(1)}(\tau)$ is calculated from a system of linear differential equations, derived from the CIM and TPM equations using the quantum regression~\cite{carmichael2002} theory. In this approach each CIM or TPM equation for a variable $\langle O(t) \rangle$ leads to a linear differential equation for the mixed time expectation value $\langle b^\dagger(t) O(t+\tau)\rangle$, where $\tau$ is the delay time~\cite{ates08a,carroll23a}. For the CIM and TPM non-lasing solutions this approach leads to a $2\times2$ and a $5\times5$ system respectively, Eqs.~(S.27) and~(S.24).  Unfortunately, this approach is not suitable for the lasing solutions. These are phase-rotation invariant and, hence, their decoherence is dictated by slow fluctuations of the phase~\cite{henry86a,protsenko21a}. Mathematically, the invariance implies that the linear equations have a zero eigenvalue and it is not \fp{straightforward} to use this theory to study their (slow) dynamics in the corresponding manifold. Therefore, we follow the approach of Ref.~\cite{chirkin11a}, where it is shown that, for systems like those considered here in which the polarization decays much faster than the other variables~\cite[Eqs.~(24) and~(30)]{chirkin11a}, $\tau_{c}$ can be estimated by the Schawlow-Townes formula, Eq.~(S.25).

The coherence time of the CIM non-lasing solution, CIM~(NL), diverges at threshold. This is due to the nature of the CIM lasing transition: the non-lasing solution changes its stability and the stable lasing solution appears with zero amplitude. As the transition point is approached the fluctuations become slower and the coherence time diverges. This is not the case of the TPM non-lasing solution, TPM~(NL), which remains always stable and with finite and small coherence time. The lasing solutions of either model, CIM~(L) and TPM~(L), have similar behaviors: their $\tau_{c}$ are much larger than the non-lasing solutions and increase with the pump. 

Because of the bistability between TPM~(L) and TPM~(NL) solutions, even above threshold a nanolaser will be in a mixture of these two states and, therefore, its coherence will be a combination of theirs. In particular, the \gd{experimentally observed} smooth increase in correlation as the bifurcation is approached~\cite{tempel11a} can be explained in terms of mixed lasing and non-lasing states with the weight of the lasing state component increasing with the pump. For illustrative purposes, it is worth noting that noise-induced bimodal optical bistability, predicted in~\cite{broggi1984transient} and observed in~\cite{lange1985study}, represents a semi-classical analogue of the quantum bistability predicted by the TPM, with the semi-classical probability density function of the interferometer transmission playing the role of the double-peaked wave-function of the quantum state.

\begin{figure}
  \centering
  \includegraphics[width=0.45\textwidth]{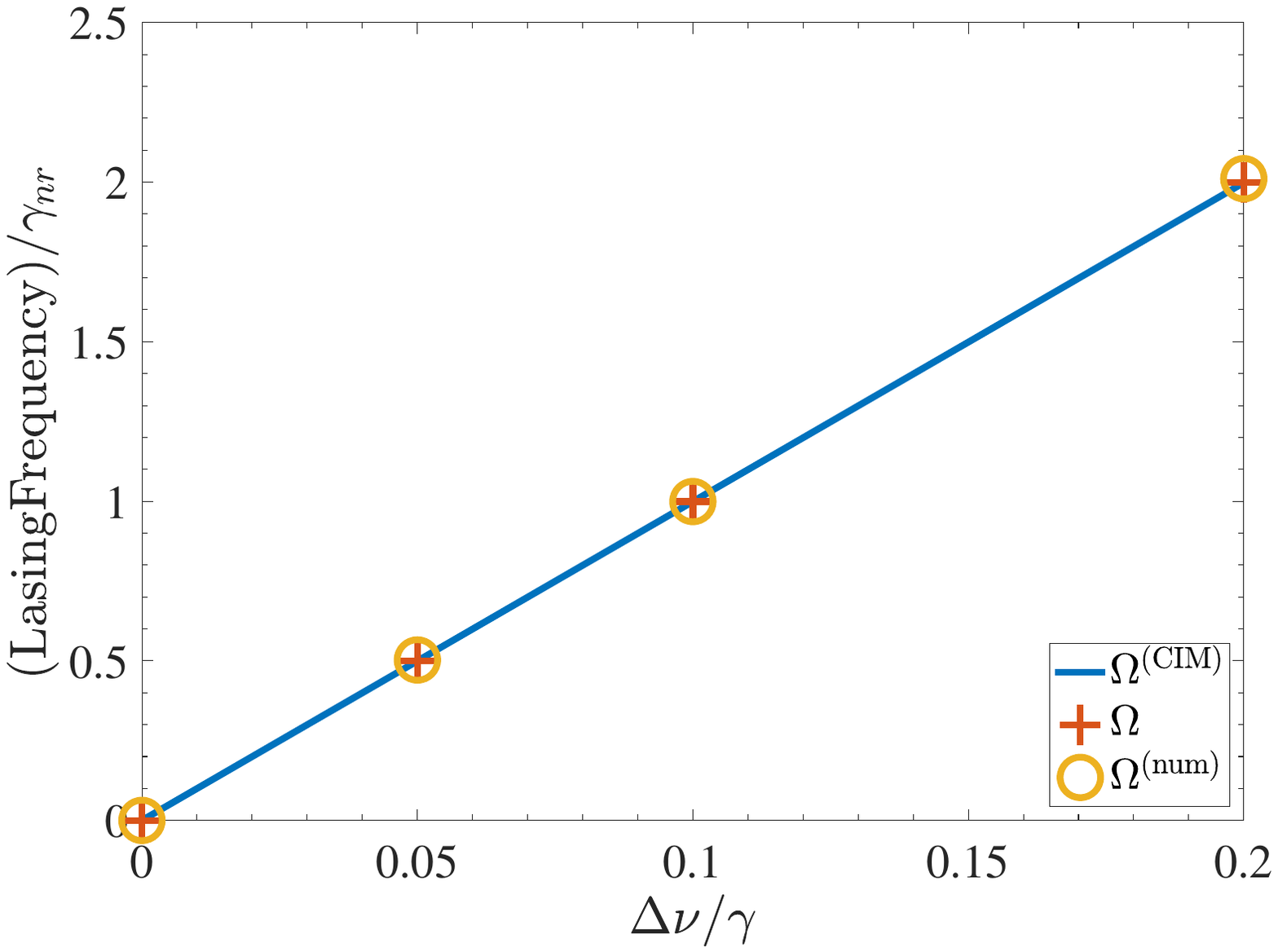} 
  \caption{Frequency vs detuning $\Delta \nu $ for the CIM and TPM (with $\mu=\{0.00,0.05\}$) lasing solutions. The solid line is the analytical CIM frequency $\Omega^{(\mathrm{CIM})}$. The $+$ markers are the CIM and TPM lasing frequencies computed using equation~\eqref{laser_frequency} averaged over $100/\gamma_{nr}$~s. The circle markers are the frequency computed by taking the dominant peak in the power spectrum of the numerical solution of the coherent field $\langle b \rangle$ averaged over the same time interval. The differences between the CIM and the two TPM simulations are not visible on the scale of the graph. All parameters as in Fig.~\ref{fig:BD} except that $r/\gamma_{nr}=8 \times 10^5$ . }
  \label{fig:Omega}
\end{figure}

We conclude the analysis of the TPM~(L) solutions with a study of their frequency. The steady state single-mode lasing frequency $\Omega$, derived analytically in section S-VI, is
\begin{equation}
  \label{laser_frequency}
  \Omega =  \nu + \gamma_{c}
  \frac{ \mathrm{Im} \left (
      \langle v^{\dagger} c  \rangle^{*} \langle b \rangle \right ) }
  {\mathrm{Re}  \left ( 
      \langle v^{\dagger} c  \rangle^{*} \langle b \rangle \right )  }. 
\end{equation}
In the case of the CIM this formula simplifies to $\Omega^{(\mathrm{CIM})} =  \nu + \gamma_c \Delta \nu/( \gamma_c + \gamma)$~\cite{carroll21a}, which appears to be also an excellent approximation of the TPM~(L) frequency, as shown in Fig.~\ref{fig:Omega}.

In conclusion, we have observed that in \gd{the TPM, i.e. a model that includes } two-particle quantum correlations, lasing appears through a \gd{Hopf} saddle-node bifurcation \fp{\gd{for both nano- and macrolasers}}. Furthermore, incoherent and coherent states are bistable, and the lasing solutions have a finite linewidth.  For nanolasers with a small number of QDs, quantum effects result in a significant departure from semi-classical theories; \fp{the same effects are present also in macro lasers, but are much harder to observe}. From the perspective of applications, this affects key properties such as the signal to noise ratio as well as the build up or collapse of coherence that are crucial in all applications where coherence is essential, like spectroscopy, and in those involving ramping up or down of the laser parameters, such as data processing and storing. The non-classical nature of the coherent states and the quantum correlations described here can have practical applications in continuous variable quantum technologies which rely on Gaussian states. Non-classical Gaussian states, present both in the CIM~\cite{carroll21b} and in the TPM, can lead to entanglement, as recently proven with two-mode Gaussian states~\cite{li23a}. The advantages of nanolasers is that they can be integrated into photonic chips leading to portable quantum optical devices~\cite{pelucchi22a}. Furthermore, the theory provides measurable quantities such as laser frequency and linewidth that allow one to identify emission processes and laser threshold.  We remark that the predicted bistability is consistent with the experimental observation of spontaneous photon burst emission at the threshold of microlasers~\cite{wang2015dynamical, wang2019exploration} for which no first-principles theory has been previously identified. \fp{Finally, \gd{preliminary} numerical simulations show that the condition of identical QDs can be relaxed without changing the scenario described in this paper even without fermionic correlations, which indicates that the effect should be observable even at higher temperatures.}


\section*{Acknowledgments}

We thank Peter Kirton and John Jeffers for many illuminating discussions
  

%

\end{document}


\title{Supplementary Material: Quantum correlations, mixed states and  bistability at the onset of lasing}%

\author{Francesco Papoff}
\email{f.papoff@strath.ac.uk}
\affiliation{Department of Physics, University
of Strathclyde,  107 Rottenrow,  Glasgow G4 0NG, UK.}
\author{Mark Anthony Carroll}
\affiliation{Department of Physics, University
of Strathclyde,  107 Rottenrow,  Glasgow G4 0NG, UK.}

\author{Gian Luca Lippi}
\affiliation{Universit\'e C\^ote d'Azur, Institut de Physique de Nice, UMR 7710 CNRS, 1361 Route des Lucioles, 06560 Valbonne, France}
\author{Gian-Luca Oppo}\affiliation{Department of Physics, University
of Strathclyde,  107 Rottenrow,  Glasgow G4 0NG, UK.}
\author{Giampaolo D'Alessandro}
\affiliation{School of Mathematical Sciences, University of Southampton, Southampton SO17 1BJ, United Kingdom}

\date{\today}


\maketitle

\section{The Heisenberg equations}

In  compact form, the Heisenberg equations for two particle operators including non radiative losses and polarization dephasing and cavity mode losses due to coupling to a Markovian bath 
in non dimensional units are
\begin{align}
  & \frac{\mathrm{d}}{\mathrm{d}t}
  \begin{bmatrix}
     O_i   \\   O_i {O}_j 
  \end{bmatrix}
  = L \begin{bmatrix}
 O_i   \\   O_i O_j 
  \end{bmatrix}
  +\begin{bmatrix}
 \mathcal{A}(O_i)   \\   \mathcal{A}(O_i O_j) 
  \end{bmatrix}
  + \begin{bmatrix} 0  \\ R O_i O_j O_k  \end{bmatrix} ,
  \label{HL-hierarchy}
\end{align}
where $L,R$ are the matrices defined in the main text and the terms $\mathcal{A}$ account for radiative decay due to non resonant modes:
\begin{align}
  \label{eq:19}
  \mathcal{A}(O_i) &= \begin{cases}
    - \sum_{q \neq 0}\left [g_{iq} b_{q} c^{\dagger}_{i} v_{i}
      + \mathrm{h.c.}\right ] & \text{if} \;\;\; O_i = c^\dagger_i c_i \\
    0 & \text{otherwise}
  \end{cases} \\*[3mm]
  \label{eq:5}
  \mathcal{A}(O_i O_j) &= \begin{cases}
    - \sum_{q \neq 0}\left [g_{iq} b_{q} c^{\dagger}_{i} v_{i}
      + \mathrm{h.c.}\right ] O_j & \text{if} \;\;\; O_i = c^\dagger_i c_i \\
    - O_i \sum_{q \neq 0}\left [g_{jq} b_{q} c^{\dagger}_{j} v_{j}
      + \mathrm{h.c.}\right ]  & \text{if} \;\;\; O_j = c^\dagger_j c_j \\     
    0 & \text{otherwise}
  \end{cases}
\end{align}
In explicit form, the equations read
{\allowdisplaybreaks
\begin{subequations}
  \label{eq:1}
  \begin{align}
    \label{eq:2}
    d_{t} b_{s}
    =&  -(\gamma_{s} +i \nu_{s}) b_{s} +\sum_{n} g^{*}_{ns} v^{\dagger}_{n} c_{n} \\
    \label{eq:3}
    d_{t} v^{\dagger}_{l} c_{l}
    =& - \left ( \gamma + i \nu_{\varepsilon_{l}} \right ) v^{\dagger}_{l} c_{l}
      + \sum_{q} g_{lq} \left [ 2b_{q} c^{\dagger}_{l} c_{l} -b_{q} \right ] \\
    d_{t} c^{\dagger}_{l} c_{l}
    =& -\gamma_{nr} c^{\dagger}_{l} c_{l}
      - \sum_{q}\left [g_{lq} b_{q} c^{\dagger}_{l} v_{l}
      + \mathrm{h.c.}\right ] \\
    d_{t} b^{\dagger}_{q} b_{q}
    =& -2 \gamma_{q} b^{\dagger}_{q} b_{q}
      + \sum_{n} \left [ g_{nq} b_{q} c^{\dagger}_{n} v_{n}
      + \mathrm{h.c.} \right ] \\
    d_{t} b_{s} c^{\dagger}_{l} v_{l}
    =& - \left ( \gamma + \gamma_{s}+ i \Delta \nu \right )
      b_{s} c^{\dagger}_{l} v_{l}
      + g^{*}_{ls} c^{\dagger}_{l} c_{l}
      -\sum_{q} g^{*}_{lq} \left (
      2 b^{\dagger}_{q} b_{s} c^{\dagger}_{l} c_{l} - b^{\dagger}_{q} b_{s} \right )
      + \sum_{n \ne l} g^{*}_{ns}c^{\dagger}_{l} v^{\dagger}_{n} c_{n} v_{l} \\
    d_{t} b_{s} c^{\dagger}_{l} c_{l}
    =& - \left (\gamma_{s} + \gamma_{nr} + i \nu_{s} \right )
      b_{s} c^{\dagger}_{l} c_{l} \
      - b_{s} \sum_{q} \left (
      g_{lq} b_{q} c^{\dagger}_{l} v_{l} + h.c. \right )
      + \sum_{n \ne l} g^{*}_{ns} v^{\dagger}_{n} c^{\dagger}_{l} c_{l} c_{n} \\
    d_{t} b_{s} b_{s}
    =&  -2 \left (\gamma_{s} +i \nu_{s} \right ) b_{s} b_{s}
      + 2 \sum_{n} g^{*}_{ns} b_{s} v^{\dagger}_{n} c_{n}   \\
    d_{t} b_{s} v^{\dagger}_{l} c_{l}
    =& - \left [ \gamma + \gamma_{s}
      + i \left (\nu_{\varepsilon_{n}} + \nu_{\varepsilon_{l}} \right )
      \right ] b_{s} v^{\dagger}_{l} c_{l}
      + g_{ls} b_{s} b_{s} \left (2 c^{\dagger}_{l} c_{l} -1\right )
      + \sum_{n \ne l} g^{*}_{ns} v^{\dagger}_{n} v^{\dagger}_{l} c_{l} c_{n} \\
    \label{eq:HE_CVvc}
    d_{t} c_{l}^{\dagger} v_{n}^{\dagger}  c_{n} v_{l}
    =& - \left [ 2\gamma
      + i \left (\nu_{\varepsilon_{n}}-\nu_{\varepsilon_{l}} \right )
      \right ] c_{l}^{\dagger} v_{n}^{\dagger} c_{n} v_{l}
      + \sum_{q} \left [
      g_{lq}^{*}\left (2b_{q}^{\dagger} v_{n}^{\dagger} c_{l}^{\dagger} c_{l} c_{n}
      - b_{q}^{\dagger} v_{n}^{\dagger} c_{n}\right ) \right . \\
    \nonumber
    & + \left . g_{nq}\left (2b_{q} c_{l}^{\dagger} c_{n}^{\dagger} c_{n} v_{l}
      - b_{q} c_{l}^{\dagger} v_{l} \right ) \right ]\\
    \label{eq:HE two particle VCcc}
    d_{t} v_{n}^{\dagger} c_{l}^{\dagger} c_{l} c_{n}
    =& -\left (\gamma + \gamma_{nr} +i\nu_{\varepsilon_{n}}\right )
      v_{n}^{\dagger} c_{l}^{\dagger} c_{l} c_{n}
      + \sum_{q}\left [
      g_{nq}\left (2 b_{q} c_{n}^{\dagger} c_{l}^{\dagger} c_{l} c_{n}
      - b_{q} c_{l}^{\dagger} c_{l} \right )
      - g_{lq}b_{q} c_{l}^{\dagger} v_{n}^{\dagger} c_{n} v_{l} \right . \\
    \nonumber
    & \left . - g_{lq}^{*} b_{q}^{\dagger} v_{n}^{\dagger} v_{l}^{\dagger} c_{l} c_{n}
      \right ] \\
    \label{eq:HE two particle CCcc}
    d_{t} c_{n}^{\dagger} c_{l}^{\dagger} c_{l} c_{n}
    =& -2\gamma_{nr}  c_{n}^{\dagger} c_{l}^{\dagger} c_{l} c_{n}
      -\sum_{q} \left [
      g_{nq}b_{q} c_{n}^{\dagger} c_{l}^{\dagger} c_{l} v_{n}
      + g_{lq}b_{q} c_{l}^{\dagger} c_{n}^{\dagger} c_{n} v_{l} + \mathrm{h.c.}
      \right ] \\
    \label{eq:HE two particle VVcc}
    d_{t} v_{n}^{\dagger} v_{l}^{\dagger} c_{l} c_{n}
    =& - \left [ 2\gamma
      + i \left (\nu_{\varepsilon_{n}} + \nu_{\varepsilon_{l}} \right ) \right ]
      v_{n}^{\dagger} v_{l}^{\dagger} c_{l}  c_{n}
      + \sum_{q}\left [
      g_{nq}\left (2b_{q} v_{l}^{\dagger} c_{n}^{\dagger} c_{n} c_{l}
      - b_{q}v_{l}^{\dagger} c_{l}\right ) \right . \\
    \nonumber
    & \left . + g_{lq}\left (2b_{q}v_{n}^{\dagger} c_{l}^{\dagger} c_{l}  c_{n}
      - b_{q}v_{n}^{\dagger} c_{n}\right )\right ], 
  \end{align}
\end{subequations}}
where $\nu_{s}$, $\nu_{\varepsilon_{n}}$ are the frequencies of the $s$-th mode and the radiative transition of the $n$-th QD and $g_{ns}$ is the light-matter coupling coefficient of the $n$-th QD with the $s$-th mode, $\gamma_{nr}$ is the population decay rate due to non radiative losses, $\gamma$ the decay rate of the polarization and $\gamma_{s}$ the decay rate of the $s$-th mode.

\section{Expectation values of three and four particle operators}

The truncation inherent in the derivation of the Coherent Incoherent Model (CIM) and Two Particle Model (TPM) equations require to expand expectation values of the product of $n$ operators in terms of lower order expectation values plus $n$-th order fluctuations, $ \langle n \rangle_\mathcal{C}$. In the truncation process these are neglected thus leading to a closed set of equations. Here we give the expansion rules for the products of order three and four.

The expectation values of three particle operators are decomposed in the sum of the three particle correlations and products of single and two particle expectation values,
\begin{equation}
  \label{eq:240306:132}
  \langle O_{i}  O_{j} O_{k} \rangle
  =  \langle O_{i}  O_{j} O_{k} \rangle_{\mathcal{C}}
  + \langle  O_{i} O_{j} O_{k}  \rangle_{\mathcal{D}},
\end{equation}
where the labels $\mathcal{C},\mathcal{D}$ stand for correlated three particle 
processes (or ``connected" in terms of Feynman diagrams) and decorrelated products 
of one and two particle processes (or ``disconnected"), with
\begin{equation}
  \label{eq:240306:133}
  \langle  O_{i} O_{j} O_{k}  \rangle_\mathcal{D}
  = \langle O_{i} \rangle \langle  O_{j} O_{k} \rangle 
  + \langle O_{j} \rangle \langle  O_{i} O_{k} \rangle
  + \langle O_{k} \rangle \langle  O_{i} O_{j} \rangle
  - 2 \langle O_{i} \rangle \langle  O_{j} \rangle \langle O_{k} \rangle.
\end{equation}
Similarly, for the expectation values of four particle operators we have
\begin{equation}
  \label{eq:240306:134}
  \langle O_{i}  O_{j} O_{k} O_{l} \rangle
  =  \langle O_{i}  O_{j} O_{k} O_{l}\rangle_{\mathcal{C}}
  + \langle  O_{i} O_{j} O_{k} O_{l} \rangle_{\mathcal{D}},
\end{equation}
where
\begin{equation}
  \label{eq:240306:135}
  \begin{split}
    \langle O_{i} O_{j} O_{k} O_{l}  \rangle_\mathcal{D}
    =& \langle O_{i} \rangle \langle O_{j} O_{k} O_{l}\rangle
    + \langle O_{j} \rangle \langle O_{i} O_{k} O_{l} \rangle
    + \langle O_{k} \rangle \langle O_{i} O_{j} O_{l} \rangle
    + \langle O_{l} \rangle \langle O_{i} O_{j} O_{k} \rangle \\
    & + \langle O_{i} O_{j} \rangle \langle O_{k} O_{l}\rangle
    + \langle O_{i} O_{k} \rangle \langle O_{j} O_{l}\rangle
    + \langle O_{i} O_{l} \rangle \langle O_{j} O_{k}\rangle \\
    & - 2 \left (
    \langle O_{i} \rangle \langle O_{j} \rangle \langle O_{k} O_{l}\rangle
    + \langle O_{i} O_{j} \rangle \langle O_{k} \rangle \langle O_{l}\rangle
    + \langle O_{i} \rangle \langle O_{k} \rangle \langle O_{j} O_{l}\rangle
    \right . \\
    & \qquad \left .
    + \langle O_{i} O_{k} \rangle \langle O_{j} \rangle \langle O_{l}\rangle
    + \langle O_{i} \rangle \langle O_{l} \rangle \langle O_{j} O_{k}\rangle
    + \langle O_{i} O_{l} \rangle \langle O_{j} \rangle \langle O_{k}\rangle
    \right ) \\
    & + 6 \langle O_{i} \rangle \langle O_{j} \rangle \langle O_{k} \rangle
    \langle O_{l} \rangle.
  \end{split}
\end{equation}

One can see that $\langle O_i \rangle$
in Eq.~\eqref{eq:240306:133} is multiplied by the same factor as 
$\langle O_i O_l\rangle$ in Eq.~\eqref{eq:240306:135}, a property that is useful in linearizing equations.
 
\section{Model equations for expectation values}

\subsection{The TPM equations in the laboratory frame}
\label{sec:tpm-equat-labor}

We derive the two particle equations for the expectation values by applying equations~\eqref{eq:240306:132} and~\eqref{eq:240306:134} to the Heisenberg equations~\eqref{eq:1} and neglecting \gd{correlations of order three or higher}.  We assume that all the quantum dots are identical so that there is no longer any need to index the fermion operators.  Moreover, the effect of spontaneous emission into the non-lasing modes is found from the adiabatic solution of the photon assisted polarization, see Eqs.~(12-13) in Ref.~\cite{gies07a}: with $q=0$ the index of the lasing mode, for the non lasing modes ($q \ne 0$) we have $\langle b_{q} \rangle =0$ and $ \langle O b_{q} \rangle_\mathcal{C} =0$ for a generic operator $O$.  The equations for the expectation values are obtained taking the expectation values of Eq.~\eqref{HL-hierarchy}, adiabatically eliminating the non resonant modes and truncating the terms at two particle level:
\begin{align}
  \frac{\mathrm{d}}{\mathrm{d}t}
  \begin{bmatrix}
    \langle O_i  \rangle  \\  \langle  O_i {O}_j  \rangle
  \end{bmatrix}
  &= L \begin{bmatrix}
    \langle O_i \rangle  \\  \langle  O_i O_j \rangle
  \end{bmatrix}
  +  \begin{bmatrix}
\langle \mathcal{A}( O_i ) \rangle \\ \langle \mathcal{A}( O_i O_j ) \rangle
  \end{bmatrix}
  + \begin{bmatrix} 0  \\ R \langle O_i O_j O_k \rangle \end{bmatrix}
  \nonumber \\
  &\simeq 
    L \begin{bmatrix}
      \langle {O}_i \rangle  \\  \langle  O_i O_j \rangle
    \end{bmatrix}
  +  \begin{bmatrix}
 \tilde{\mathcal{A}}( \langle O_i \rangle ) \\  \tilde{\mathcal{A}}( \langle O_i \rangle \langle O_j \rangle) 
  \end{bmatrix}
  + \begin{bmatrix} 0  \\ R  \langle  O_i O_j O_k  \rangle_\mathcal{D} \end{bmatrix}.
  \label{EV-hierarchy}
\end{align}
Including incoherent pumping due to population transfer from wetting layer states in a similar way to the decay from non lasing modes, the adiabatic elimination is done as follows: 
\begin{align}
\tilde{\mathcal{A}}( \langle O_i \rangle ) = \begin{cases}
 - \gamma_{nl} \langle c^\dagger_i c_i  \rangle + r(1 - \langle c^\dagger_i c_i \rangle ) & \text{if} \;\;\; O_i = c^\dagger_i c_i \\
 0 & \text{otherwise}
\end{cases} 
\end{align}
\begin{align}
 \tilde{\mathcal{A}}( \langle O_i \rangle \langle O_j \rangle) = \begin{cases}
 (- \gamma_{nl} \langle c^\dagger_i c_i  \rangle + r(1 - \langle c^\dagger_i c_i \rangle) \langle O_j \rangle & \text{if} \;\;\; O_i = c^\dagger_i c_i \\
   \langle O_i \rangle   (- \gamma_{nl} \langle c^\dagger_j c_j  \rangle + r(1 - \langle c^\dagger_j c_j \rangle) & \text{if} \;\;\; O_j = c^\dagger_j c_j \\     
 0 & \text{otherwise}
\end{cases}
\end{align}
\fp{$\gamma_{nl}$ is the decay rate due to the non lasing modes, $r$ is the pump rate per emitter.} We obtain the two particle model (TPM) equations:
{\allowdisplaybreaks
  \begin{subequations}
    \label{eq:4}
    \begin{align}
      \label{eq:b}
      d_{t} \langle   b  \rangle
      =& - ( \gamma_c +i \nu)  \langle b  \rangle
        + N g^{*} \langle v^{\dagger} c  \rangle,\\
      \label{eq:Vc}
      d_{t} \langle v^{\dagger} c  \rangle
      =& - \left ( \gamma +i \nu_\varepsilon \right )
        \langle v^{\dagger} c  \rangle
        + g \left ( 2 \langle   b c^{\dagger} c \rangle
        - \langle   b  \rangle \right ).  \\
      \label{eq:Cc}
      d_{t} \langle  c^{\dagger} c \rangle
      =& - \gamma_{nr}\langle c^{\dagger} c \rangle
        - \left ( g \langle  b c^{\dagger} v \rangle + \mathrm{h.c.} \right )
        + \left [ - \gamma_{nl} \langle  c^{\dagger} c \rangle
        + r \left (1-\langle  c^{\dagger} c \rangle \right ) \right ] , \\
      \label{eq:Bb}
      d_{t} \langle  b^{\dagger} b \rangle
      =& -2 \gamma_c \langle  b^{\dagger} b \rangle
        + N \left (g \langle  b c^{\dagger} v \rangle
        + \mathrm{h.c.} \right ), \\
      \label{eq_{s}f:bCv}
      d_{t} \langle   b c^{\dagger} v \rangle
      =& - \left (\gamma + \gamma_c + i \Delta \nu \right )
        \langle   b c^{\dagger} v \rangle
        + g^{*} \left [ \langle  c^{\dagger} c \rangle
         + 2 \left \langle 
         b^{\dagger} b  c^{\dagger} c  \right \rangle_\mathcal{D}
        - \langle   b^{\dagger} b   \rangle \right ]
        +  (N-1) g^{*} \langle c^{\dagger} v^{\dagger} c v \rangle \\
      \label{eq_{s}f:bCc}
      d_{t} \langle b c^{\dagger} c  \rangle
      =& - \left ( \gamma_c + \gamma_{nr} + i \nu \right )
        \langle b c^{\dagger} c  \rangle
        -g \left \langle  b b c^{\dagger} v  \right \rangle_\mathcal{D}
         -g^{*} \left \langle 
         b^{\dagger}  b v^{\dagger} c  \right \rangle_\mathcal{D}
         + (N-1) g^{*} \langle c^{\dagger} v^{\dagger} c c \rangle  \\
      \nonumber
      & + \langle b \rangle \left [-\gamma_{nl} \langle c^{\dagger} c \rangle
        + r \left (1-\langle  c^{\dagger} c \rangle \right )
        \right ] \\
      \label{eq:bb}
      d_{t} \langle b b \rangle
      =& - 2 \left ( \gamma_c + i  \nu \right ) \langle b b \rangle
        + 2 N g^{*} \langle b v^{\dagger} c  \rangle \\
      \label{eq_{s}f:bVc}
      d_{t} \langle b v^{\dagger} c  \rangle
      =& - \left [ \gamma_c + \gamma
        + i \left ( \nu + \nu_{\varepsilon} \right ) \right ]
        \langle b v^{\dagger} c  \rangle
         + g \left [
         2 \left \langle  b b   c^{\dagger} c  \right \rangle_\mathcal{D}
        - \langle b b \rangle \right ]
        + (N-1) g^{*}  \langle  v^{\dagger} v^{\dagger} c c \rangle \\
      d_{t} \langle c^{\dagger} v^{\dagger} c v \rangle
      =& -2 \gamma(1+\mu) \langle c^{\dagger} v^{\dagger} c v\rangle
         + g^{*}\left [
         2 \left \langle 
         b^{\dagger} v^{\dagger} c^{\dagger} c c  \right \rangle_\mathcal{D}
         - \left \langle b^{\dagger} v^{\dagger} c \right \rangle \right ]  \\
      \nonumber
      & + g \left [
        2 \left \langle 
        b c^{\dagger} c^{\dagger} c v  \right \rangle_\mathcal{D}
        - \langle b c^{\dagger} v \rangle \right ] \\
      d_{t} \langle v^{\dagger} c^{\dagger} c c \rangle
      =& -\left [ \gamma(1+\mu) +\gamma_{nr}+i\nu_\varepsilon \right ]
         \langle v^{\dagger} c^{\dagger} c c \rangle
         + g \left [ 2 \left \langle 
         b c^{\dagger} c^{\dagger} c c  \right \rangle_\mathcal{D}
         - \langle b c^{\dagger} c\rangle \right ] \\
      \nonumber
       & - g \left \langle 
         b c^{\dagger} v^{\dagger} c  v  \right \rangle_\mathcal{D}
        - g^{*} \left \langle  b^{\dagger} v^{\dagger} v^{\dagger} c  c
         \right \rangle_\mathcal{D}
        + \langle v^{\dagger} c \rangle \left [
        - \gamma_{nl} \langle c^{\dagger} c \rangle
        + r \left (1-\langle c^{\dagger} c\rangle \right ) \right ],\\
      d_{t} \langle c^{\dagger} c^{\dagger} c c\rangle
      =& -2\gamma_{nr}\langle c^{\dagger} c^{\dagger} c c\rangle
         - \left [
         2 g \left \langle 
         b c^{\dagger} c^{\dagger} c  v \right \rangle_\mathcal{D}
         + \mathrm{h.c.} \right ]  \\
      \nonumber
      & + 2 \langle c^{\dagger} c \rangle
        \left [- \gamma_{nl}\langle c^{\dagger} c \rangle
        +  r \left (1-\langle c^{\dagger} c \rangle \right ) \right ], \\
      d_{t}\langle v^{\dagger} v^{\dagger} c c\rangle
      =& - 2\left [\gamma(1+\mu) + i \nu_\varepsilon \right ]
         \langle v^{\dagger} v^{\dagger} c c\rangle
         + 2 g\left [2 \left \langle 
         b v^{\dagger} c^{\dagger} c c  \right \rangle_\mathcal{D} 
         - \langle b v^{\dagger} c\rangle \right ] ,
    \end{align} 
  \end{subequations}
}
where $\gamma \mu$, with $\mu \ge 0$, is the dephasing rate due to phonon scattering, $\nu$, $\nu_\varepsilon$ are the frequencies of the laser mode and of the emitters' transition, respectively, and $\Delta \nu = \nu - \nu_{\varepsilon}$.

\subsection{The TPM equations in the rotating reference frame}
\label{sec:thetpm-equat-rotat:101123}

With a slight abuse of notation we indicate with an $H$ subscript the expectation values in the Heisenberg picture in the laboratory frame, equations~\eqref{eq:4}. The expectation values in the rotating frames that are different from those in the Heisenberg picture are:
\begin{subequations}
  \label{eq:240306:56}
  \begin{align}
    \label{eq:240306:57}
    & \langle b \rangle = \langle b \rangle_{H} e^{i\nu t}, \\
    \label{eq:240306:58}
    & \langle v^{\dagger} c \rangle
      = \langle v^{\dagger} c \rangle_{H} e^{i\nu t}, \\
    \label{eq:240306:59}
    & \langle b c^{\dagger} c \rangle
      = \langle b c^{\dagger} c \rangle_{H} e^{i\nu t}, \\
    \label{eq:240306:62}
    & \langle v^{\dagger} c^{\dagger} cc\rangle
      = \langle v^{\dagger} c^{\dagger} cc\rangle_{H} e^{i\nu t}, \\
    \label{eq:240306:60}
    & \langle b v^{\dagger} c \rangle
      = \langle b v^{\dagger} c \rangle_{H} e^{i 2 \nu t}, \\
    \label{eq:240306:61}
    & \langle bb \rangle = \langle bb \rangle_{H} e^{i 2 \nu t} \\
    \label{eq:240306:63}
    & \langle v^{\dagger} v^{\dagger} c c  \rangle
      = \langle v^{\dagger} v^{\dagger} c c  \rangle_{H} e^{i 2 \nu t}.
  \end{align}
\end{subequations}
The equations for these expectation values in  the rotating frame are
{\allowdisplaybreaks
  \begin{subequations}
    \label{eq:6}
    \begin{align}
      \label{eq:7}
      d_{t} \langle   b  \rangle
      =& - \gamma_c \langle b  \rangle
         + N g^{*} \langle v^{\dagger} c  \rangle,\\
      \label{eq:8}
      d_{t} \langle v^{\dagger} c  \rangle
      =& - \left ( \gamma - i \Delta \nu \right )
        \langle v^{\dagger} c  \rangle
        + g \left ( 2 \langle   b c^{\dagger} c \rangle
         - \langle   b  \rangle \right ).  \\
      \label{eq:9}
      d_{t} \langle  c^{\dagger} c \rangle
      =& - \gamma_{nr}\langle c^{\dagger} c \rangle
        - \left ( g \langle  b c^{\dagger} v \rangle + \mathrm{h.c.} \right )
        + \left [ - \gamma_{nl} \langle  c^{\dagger} c \rangle
        + r \left (1-\langle  c^{\dagger} c \rangle \right ) \right ] , \\
      \label{eq:10}
      d_{t} \langle  b^{\dagger} b \rangle
      =& -2 \gamma_c \langle  b^{\dagger} b \rangle
        + N \left (g \langle  b c^{\dagger} v \rangle
        + \mathrm{h.c.} \right ), \\
      \label{eq:11}
      d_{t} \langle   b c^{\dagger} v \rangle
      =& - \left (\gamma + \gamma_c + i \Delta \nu \right )
        \langle   b c^{\dagger} v \rangle
        + g^{*} \left [ \langle  c^{\dagger} c \rangle
         + 2 \left \langle 
         b^{\dagger} b  c^{\dagger} c  \right \rangle_\mathcal{D}
        - \langle   b^{\dagger} b   \rangle \right ]
        +  (N-1) g^{*} \langle c^{\dagger} v^{\dagger} c v \rangle \\
      d_{t} \langle b c^{\dagger} c  \rangle
      =& - \left ( \gamma_c + \gamma_{nr} \right )
        \langle b c^{\dagger} c  \rangle
        -g \left \langle  b b c^{\dagger} v  \right \rangle_\mathcal{D}
         -g^{*} \left \langle 
         b^{\dagger}  b v^{\dagger} c \right \rangle_\mathcal{D}
         + (N-1) g^{*} \langle c^{\dagger} v^{\dagger} c c \rangle  \\
      \nonumber
      & + \langle b \rangle \left [-\gamma_{nl} \langle c^{\dagger} c \rangle
        + r \left (1-\langle  c^{\dagger} c \rangle \right )
        \right ] \\
      d_{t} \langle b b \rangle
      =& - 2 \gamma_c \langle b b \rangle
        + 2 N g^{*} \langle b v^{\dagger} c  \rangle \\
      d_{t} \langle b v^{\dagger} c  \rangle
      =& - \left [ \gamma_c + \gamma - i \Delta \nu  \right ]
        \langle b v^{\dagger} c  \rangle
         + g \left [
         2 \left \langle  b b   c^{\dagger} c \right \rangle_\mathcal{D}
        - \langle b b \rangle \right ]
        + (N-1) g^{*}  \langle  v^{\dagger} v^{\dagger} c c \rangle \\
      d_{t} \langle c^{\dagger} v^{\dagger} c v \rangle
      =& -2 \gamma(1+\mu) \langle c^{\dagger} v^{\dagger} c v\rangle
         + g^{*}\left [
         2 \left \langle 
         b^{\dagger} v^{\dagger} c^{\dagger} c c  \right \rangle_\mathcal{D}
         - \left \langle b^{\dagger} v^{\dagger} c \right \rangle \right ]  \\
      \nonumber
      & + g \left [
        2 \left \langle 
        b c^{\dagger} c^{\dagger} c v  \right \rangle_\mathcal{D}
        - \langle b c^{\dagger} v \rangle \right ] \\
      d_{t} \langle v^{\dagger} c^{\dagger} c c \rangle
      =& -\left [ \gamma(1+\mu) +\gamma_{nr}-i\Delta \nu \right ]
         \langle v^{\dagger} c^{\dagger} c c \rangle
         + g \left [ 2 \left \langle 
         b c^{\dagger} c^{\dagger} c c  \right \rangle_\mathcal{D}
         - \langle b c^{\dagger} c\rangle \right ] \\
      \nonumber
       & - g \left \langle 
         b c^{\dagger} v^{\dagger} c  v  \right \rangle_\mathcal{D}
        - g^{*} \left \langle  b^{\dagger} v^{\dagger} v^{\dagger} c  c
         \right \rangle_\mathcal{D}
        + \langle v^{\dagger} c \rangle \left [
        - \gamma_{nl} \langle c^{\dagger} c \rangle
        + r \left (1-\langle c^{\dagger} c\rangle \right ) \right ],\\
      d_{t} \langle c^{\dagger} c^{\dagger} c c\rangle
      =& -2\gamma_{nr}\langle c^{\dagger} c^{\dagger} c c\rangle
         - \left [
         2 g \left \langle 
         b c^{\dagger} c^{\dagger} c  v  \right \rangle_\mathcal{D}
         + \mathrm{h.c.} \right ]  \\
      \nonumber
      & + 2 \langle c^{\dagger} c \rangle
        \left [- \gamma_{nl}\langle c^{\dagger} c \rangle
        +  r \left (1-\langle c^{\dagger} c \rangle \right ) \right ], \\
      d_{t}\langle v^{\dagger} v^{\dagger} c c\rangle
      =& - 2\left [\gamma (1+\mu) - i \Delta \nu \right ]
         \langle v^{\dagger} v^{\dagger} c c\rangle
         + 2 g\left [2 \left \langle 
         b v^{\dagger} c^{\dagger} c c \right \rangle_\mathcal{D} 
         - \langle b v^{\dagger} c\rangle \right ] ,
    \end{align} 
  \end{subequations}
}

\subsection{The CIM equations in the rotating reference frame}

The CIM equations are~\eqref{eq:7} to~\eqref{eq:11} suitably
truncated:
\begin{subequations}
  \label{eq:13}
  \begin{align}
    \label{eq:14}
    d_{t} \langle   b  \rangle
    =& - \gamma_c \langle b  \rangle
       + N g^{*} \langle v^{\dagger} c  \rangle,\\
    \label{eq:15}
    d_{t} \langle v^{\dagger} c  \rangle
    =& - \left ( \gamma - i \Delta \nu \right )
       \langle v^{\dagger} c  \rangle
       + g \langle   b  \rangle
       \left ( 2 \langle c^{\dagger} c \rangle - 1\right ).  \\
    \label{eq:16}
    d_{t} \langle  c^{\dagger} c \rangle
    =& - \left ( \gamma_{nr} + \gamma_{nl} \right ) \langle c^{\dagger} c \rangle
       - \left ( g \langle  b c^{\dagger} v \rangle + \mathrm{h.c.} \right )
       + r \left (1-\langle  c^{\dagger} c \rangle \right ) , \\
    \label{eq:17}
    d_{t} \langle  b^{\dagger} b \rangle
    =& -2 \gamma_c \langle  b^{\dagger} b \rangle
       + N \left (g \langle  b c^{\dagger} v \rangle
       + \mathrm{h.c.} \right ), \\
    \label{eq:18}
    d_{t} \langle   b c^{\dagger} v \rangle
    =& - \left (\gamma + \gamma_c + i \Delta \nu \right )
       \langle   b c^{\dagger} v \rangle
       + g^{*} \left [ \langle  c^{\dagger} c \rangle
       + \langle b^{\dagger} b \rangle
       \left ( 2 \left \langle c^{\dagger} c \right \rangle - 1 \right ) 
       \right ]
       + (N-1) g^{*} \langle c^{\dagger} v \rangle
       \langle v^{\dagger} c \rangle \, .
  \end{align}
\end{subequations}

\section{First order correlation $g^{(1)}(\tau)$}

The first order correlation function \fp{for coherent and incoherent solutions} is defined as~\gd{\cite[Eq.~(2.4-24)]{MandelWolf95}}
\begin{equation}
  \label{eq:20}
  g^{(1)}(\tau) \equiv
  \frac{\langle b^\dagger(t) b(t +\tau) \rangle_\mathcal{C} }
  { \langle b^\dagger(t) b(t) \rangle_\mathcal{C}  },
\end{equation}
where $t$ is large enough for the system to have reached a steady state.  Below the laser threshold the correlation $\langle b^\dagger(t) b(t +\tau) \rangle$ vanishes as $\tau \rightarrow \infty$ and we find that photon correlations are negligible for $\tau>\tau_{c}$, where
\begin{equation}
  \label{eq:21}
  \tau_{c}=  2 \int_0^\infty |g^{(1)}(\tau)|^2 d \tau.
\end{equation}
$g^{(1)}(\tau)$ can be found using the quantum regression formula~\cite[Eqs.(1.105,1.107)]{carmichael2002},
\begin{align}
  \label{eq:22}
  & d_t \langle O_i  \rangle = \mathcal{L} \langle O_i  \rangle, \\
   \label{QRlin}
  & d_{\tau} \langle \tilde{b}^{\dagger} O_i  \rangle  =
    \mathcal{L} \langle \tilde{b}^{\dagger}  O_i  \rangle,
\end{align}
where $ \tilde{b}^{\dagger} \equiv b^{\dagger}(t)$, $O$ is an arbitrary operator that depends on $\tau$ and $\mathcal{L} $ is a matrix.

For solutions with constant field amplitudes, i.e. zero detuning, equations~\eqref{QRlin} have the form
\begin{equation}
  \label{eq:240306:38}
  \dot{\boldsymbol{x}} = A \boldsymbol{x} + \boldsymbol{q},
\end{equation}
where $A$ is a constant matrix whose entries are single time expectation values up to order $M_{1}-1$, $\boldsymbol{q}$ is a combination of single time expectation values for coherent solutions,\fp{with $\boldsymbol{q}=0$ 
($\boldsymbol{q}\neq 0$) for incoherent (coherent) solutions}, and $\boldsymbol{x}$ are \gd{two-time} expectation values. In particular, the first component of $\boldsymbol{x}$ is the first order correlation function, $g^{(1)}(\tau)$. 
%
%

Therefore 
we have computed the first order correlation function in the following way:

\begin{itemize}
\item \emph{Non-lasing TPM solution} - In the non-lasing solution the expectation value of all fast variables are zero, i.e.
  \begin{equation}
    \label{eq:240306:131}
    \langle b \rangle = \langle b b \rangle = \langle v^{\dagger} c \rangle
    = \langle b c^{\dagger} c \rangle = \langle b v^{\dagger} c \rangle
    = \langle v^{\dagger} c^{\dagger} c c \rangle 
    = \langle v^{\dagger} v^{\dagger}  c c\rangle 
    = 0.
  \end{equation}
  In this case, the remaining equations~\eqref{eq:240306:38}, block-decouple and we can write the vector $\boldsymbol{x}$ that contains the correlation function as
  \begin{equation}
    \label{eq:240306:139}
    \boldsymbol{x} = \left [\langle \tilde{b}^{\dagger}  b  \rangle,
      \langle \tilde{b}^{\dagger} v^{\dagger} c  \rangle,
      \langle \tilde{b}^{\dagger}  b c^{\dagger} c  \rangle,
      \langle \tilde{b}^{\dagger} v^{\dagger} c^{\dagger} c c \rangle
    \right ] 
  \end{equation}
  and the coefficient matrix $A$ as
  \begin{equation}
    \label{eq:240306:140}
    A =
    \begin{pmatrix}
      -\gamma_{c} & N g & 0 & 0 \\
      -g & -\gamma & 2 g & 0 \\*[2mm]
      \substack{- 3 g \langle b^{\dagger} v^{\dagger} c \rangle \\
        -\gamma_{nl} \langle c^{\dagger} c \rangle
        + r  \left (1-\langle  c^{\dagger} c \rangle\right )}
      & -g \langle b^{\dagger} b \rangle
      & - ( \gamma_{c} + \gamma_{nr}) & (N-1) g \\*[4mm]
      \substack{2 g\left (\langle c^{\dagger} c^{\dagger} c c \rangle
          - 2 \langle c^{\dagger} c \rangle \langle c^{\dagger} c \rangle 
        \right )\\ - g \langle v^{\dagger} c^{\dagger} v c \rangle}
      & \substack{-g \langle b c^{\dagger} v \rangle
        -2 g \langle b^{\dagger} v^{\dagger} c \rangle\\
        - \gamma_{nl} \langle c^{\dagger} c \rangle 
        + r\left (1-\langle c^{\dagger} c\rangle \right )}
      & 4 g \langle c^{\dagger} c \rangle - g
      & -\left [ \gamma(1+\mu) +\gamma_{nr} \right ]
    \end{pmatrix} \, .
  \end{equation}
  The vector $\boldsymbol{q}$ is zero.
\item \emph{Lasing TPM solution} - In this case, as a result of the phase invariance of the equations, the coefficient matrix $A$ has an eigenvalue that is equal to zero for all values of the pump, while the remaining eigenvalues are all negative above threshold.  Therefore, phase perturbations have a longer lifetime than all the other perturbations~\cite{henry86a,protsenko21a} and give the dominant contribution to the coherence time except in a small range of pump values near the laser threshold, where others eigenvalues of $A$ are also close to zero.  \gd{Estimating these effects for the lasing solution, may require to keep track of higher order correlation than the TPM currently does, a very complex task as the number of terms in the model increases exponentially with the order of the non-zero correlations. }
\fp{Here instead we take advantage of the fact that, sufficiently away from threshold,} it is possible to formally derive the Schawlow-Townes formula for the CIM coherence time~\cite{chirkin11a}, $\tau_c \sim \Gamma^{-1}$, with $\Gamma$ the laser line width~\cite{milonni2010} 
  \begin{align}
    \Gamma = \left(\frac{\gamma_c}{\gamma_{nr}}\right)
    \frac{1}{4 |\langle b \rangle|^2 } 
    \frac{\langle c^\dagger c \rangle}{2 \langle  c^\dagger c \rangle- 1 }
    \left( \frac{\gamma}{\gamma+\gamma_c/2} \right)^2 \label{QD_MST}.
  \end{align}
  We apply this formula also to estimate the coherence time of the TPM lasing solution.
  
\item \emph{Non-lasing CIM solution} - The non-lasing CIM solution has $\langle b \rangle = \langle v^{\dagger} c \rangle=0$. As in the non-lasing TPM case the remaining equations decouple. The vector $\boldsymbol{q}=0$, the vector $\boldsymbol{x}$ is
  \begin{equation}
    \label{eq:240306:52}
    \boldsymbol{x} = \left [\langle \tilde{b}^{\dagger} b \rangle ,
      \langle \tilde{b}^{\dagger} v^{\dagger}c \rangle \right ] ,
  \end{equation}
  and the coefficient matrix is 
  \begin{equation}
    \label{eq:240306:53}
    A = \begin{pmatrix}
      -\gamma_{c} & g N \\
      g \left ( 2 \langle c^{\dagger} c \rangle - 1 \right ) & -\gamma
    \end{pmatrix} \, .
  \end{equation}
\end{itemize}

\section{Analytical approximation of the TPM laser threshold}

The non-lasing solution of the TPM model can be found analytically by setting all the derivatives
of the incoherent variables and the coherent variables to zero. Writing all incoherent variables as functions of $\langle c^\dagger c \rangle$, we find \gd{the following relation between the pump parameter $r$ and the expectation value of the population inversion, $\langle c^\dagger c \rangle$,}
\begin{equation}
  \label{r_tpm}
  r= \left ( \Gamma_n + \frac{\Gamma_g}{K} \right )
  \frac{\langle c^\dagger c \rangle}{1-\langle c^\dagger c \rangle} ,
\end{equation}
where $\Gamma_n = \gamma_{nr} +\gamma_{nl}$, $\Gamma_g = 2 g^2 (\gamma + \gamma_c)/[(\gamma + \gamma_c)^2 + \Delta \nu^2]$ and $K = 1-\Gamma_g (\frac{N}{2 \gamma_c} + \frac{N-1}{2 \gamma (1+\mu)})(2 \langle c^\dagger c \rangle - 1)$.  The analytical estimate of the pump threshold, $r_{th}$, for the TPM model given in the main text is found by inserting in Eq.~\eqref{r_tpm} the threshold value of the CIM model~\cite[Eq.~(6)]{carroll21a}
\begin{equation}
  \langle c^\dagger c \rangle_{th} =  \frac{1}{2}
  \left \{ 1+  \frac{\gamma_c \gamma}{N g^2}
    \left [ 1 +(\frac{\Delta \nu}{\gamma_c + \gamma})^2\right ] \right \}.
\end{equation}
The accuracy of this analytical approximation is due to the fact that $\langle c^\dagger c \rangle$ is almost the same in all models and for both lasing and non-lasing solution.

\section{Laser frequency}
To find the laser frequency for a single mode laser, we start from 
\begin{equation}
  \langle b \rangle = |\langle b \rangle | \exp{(-i\phi(t))},
\end{equation}
where $|\langle b \rangle| = (\langle b \rangle \langle b^{\dagger} \rangle)^{1/2} $, from which we find the phase and its time derivative
\begin{align}
  & \phi = - \frac{1}{2i} \log \left (
    \frac{\langle b \rangle}{\langle b\dagger \rangle} \right ) \\
  & d_{t} \phi = - \frac{1}{2i} 
    \frac{(d_{t} \langle b \rangle)\langle b^{\dagger} \rangle
    - (d_{t} \langle b^{\dagger} \rangle)\langle b \rangle}
    {|\langle b \rangle |^2}.\label{phase_der}
\end{align}
Combining Eq.~\eqref{phase_der} with the steady state lasing condition from equation~\eqref{eq:7},
\begin{equation}
  \label{eq:12}
  d_{t} \langle b \rangle \langle b^{\dagger} \rangle
  = -2 \gamma_c |\langle b \rangle|^{2}
  + 2 N \mathrm{Re} \left (
    g \langle v^{\dagger} c  \rangle^{*} \langle b \rangle
  \right ) =0,
\end{equation}
we find that the  frequency of a single mode steady state laser is
\begin{equation}
  \label{laser_frequency}
  \Omega =  \nu + \gamma_c
  \frac{ \mathrm{Im} \left (
      \langle v^{\dagger} c  \rangle^{*} \langle b \rangle \right ) }
  {\mathrm{Re}  \left ( 
      \langle v^{\dagger} c  \rangle^{*} \langle b \rangle \right )  },
\end{equation}
where we have taken advantage of the phase invariance of the equations to set $g$ real and eliminate it.



\section{Parameters}

\begin{center}
\captionof{table}{\label{TofNV}Numerical values of the main parameters used in the simulations.}
\begin{tabular}{||c | c ||}\hline\hline
$g$ & $7 \times 10^{10} s^{-1}$ \\\hline
$\gamma_{nr}$ & $10^9 s^{-1}$ \\\hline
$\gamma$ & $10^{13} s^{-1}$\\\hline
$\gamma_c$ & $10^{10} s^{-1}$ \\\hline

\end{tabular}

\end{center}

The \gd{values of the} radiation-matter coupling, $g$, the non-radiative loss constant, $\gamma_{nr}$, and the polarization's relaxation rate, $\gamma$, \gd{listed in table~\ref{TofNV}} are consistent with values used in~\cite{chow14a}.  The cavity loss constant, $\gamma_c$, is chosen in agreement with a cavity quality factor $Q$ of the order of $10^4$~\cite{wiersig2009direct}.  The range in which we have chosen the number of Quantum Dots $N$ reflects experimental realizations~\cite{Kreinberg2017a}.  The value for fraction of spontaneous emission coupled into the lasing mode, $\beta$, is self-consistently calculated from the number of non-lasing modes ($\gamma_{nl}$) and the lasing mode~\cite{gies07a}. Notice that in the main article all rates are normalized to $\gamma_{nr}$.


%